\documentclass[11pt]{article}
\usepackage{authblk}
\usepackage[utf8]{inputenc}
\usepackage[T1]{fontenc}
\usepackage{graphicx}
\usepackage{amsmath}
\usepackage{amssymb}
\usepackage{hyperref}
\usepackage[margin=25mm]{geometry}
\usepackage{parskip}

\usepackage[table]{xcolor}
\usepackage{pgf}

\usepackage{colortbl}
\usepackage{multirow}
\usepackage{diagbox}
\usepackage{longtable}

\newcommand{\mythead}[1]{\begin{tabular}{@{}c@{}}#1\end{tabular}}
\definecolor{low}{HTML}{76f013}  
\definecolor{middle}{HTML}{fce303}
\definecolor{high}{HTML}{ec462e}  

\newcommand*{\opacity}{80}
\newcommand{\gradient}[4]{
  \ifdimcomp{#1pt}{<}{#2 pt}{
    \cellcolor{low!\opacity}#1
  }{
    \ifdimcomp{#1pt}{<}{#3 pt} {
      \pgfmathparse{int(round(100*(#1/(#3-#2))-(#2*(100/(#3-#2)))))}
      \xdef\tempalow{\pgfmathresult}
      \cellcolor{low!\opacity}#1
    }{
      \ifdimcomp{#1pt}{<}{#4 pt}{
        \pgfmathparse{int(round(100*(#1/(#4-#3))-(#3*(100/(#4-#3)))))}
        \xdef\tempahigh{\pgfmathresult}
        \cellcolor{middle!\opacity}#1
      }{
        \cellcolor{high!\opacity}#1
      }
    }
  }
}

\author[1]{Kirill Morkrov}
\author[1]{Alexander Smirnov}
\author[2]{Mao Zeng}
\affil[1]{Research Computing Center, Moscow State University, 119992 Moscow, Russia}
\affil[2]{Higgs Centre for Theoretical Physics, University of Edinburgh,
James Clerk Maxwell Building, Peter Guthrie Tait Road, Edinburgh, EH9 3FD,
United Kingdom}
\affil[ ]{\{kmokrov@mail.ru, asmirnov80@gmail.com, mao.zeng@ed.ac.uk\}}
\date{\today}
\title{Rational Function Simplification for Integration-by-Parts Reduction and Beyond}

\begin{document}
\maketitle

\begin{abstract}
  We present FUEL (Fractional Universal Evaluation Library), a C++ library for performing rational function arithmetic with a flexible choice of third-party computer algebra systems as simplifiers. FUEL is an outgrowth of a C++ interface to Fermat which was originally part of the FIRE code for integration-by-parts (IBP) reduction for Feynman integrals, now promoted to be a standalone library and with access to simplifiers other than Fermat. We compare the performance of various simplifiers for standalone benchmark problems as well as IBP reduction runs with FIRE. A speedup of more than 10 times is achieved for an example IBP problem related to off-shell three-particle form factors in $\mathcal N=4$ super-Yang-Mills theory.
\end{abstract}


\section{Introduction}
Many problems of high energy physics and quantum field theory are difficult to solve without using a computer. An example of such a problem is the calculation of Feynman integrals in complicated scattering amplitudes and correlation functions. For cutting-edge problems involving a huge number of Feynman integrals, the standard calculation workflow consists of two stages: integration-by-parts (IBP) reduction \cite{algorithm_to_calculete_feynman_integral, Laporta:2000dsw} to so-called master integrals and finding the values of these master integrals. The problem of IBP reduction with the \emph{Laporta algorithm} \cite{Laporta:2000dsw} can be viewed as a problem of solving a huge system of sparse linear equations with polynomial coefficients. The coefficients generally become rational functions, i.e.\ fractions of polynomials, when solving the linear system via (variants of) Gaussian elimination.

Because of the complex nature of the coefficients, they need to be stored in a special form, and most importantly, the coefficients need to be periodically simplified when solving the linear system. The simplifications include e.g.\ collecting similar terms in polynomials, writing sums of fractions as a single fraction with a common denominator, and simplifying the numerator and denominator by computing polynomial greatest common denominator (GCD).
Without the simplifications, arithmetic operations on the coefficients will take more and more time, and their storage will require more and more memory, eventually making performance unacceptable.

In this paper we consider programs (either standalone programs or libraries), called \textit{simplifiers}, which are used to perform all necessary simplifying transformations of rational function coefficients. The list of simplifiers considered in this paper is: CoCoA \cite{CoCoA, CoCoALib}, Fermat \cite{Fermat}, FORM \cite{FORM}, GiNaC \cite{GiNaC}, Macaulay2 \cite{Macaulay2}, Maple \cite{Maple}, Maxima \cite{Maxima}, Nemo \cite{Nemo}, PARI / GP \cite{PARI2}, Symbolica \cite{Symbolica}, and Wolfram Mathematica \cite{Math}. These programs are compared for three different sets of input data: a large set of rational functions in one variable, a large set of rational functions in three variables, and a set of a few dozen huge rational functions whose lengths range from tens of thousands to several hundred thousand characters when printed. The main performance indicators for comparison are the time spent on simplification and the amount of memory needed.

At the end of the 20th century, the task of IBP reduction was done manually. Later, computer programs appeared that automated and speeded up this process. Some of the publicly available general-purpose programs are: FIRE \cite{Smirnov:2013dia, Smirnov:2014hma, FIRE6}, AIR \cite{AIR}, Reduze \cite{Reduze}, LiteRed \cite{LiteRed}, and Kira \cite{Maierhofer:2017gsa, Kira, Klappert:2020nbg}. The public version of FIRE was first published in 2014 and has been used by the scientific community to perform cutting-edge calculations, e.g.\ in Refs.\ \cite{FIRE_usage_1, FIRE_usage_2, FIRE_usage_3}. Initially, Fermat was the only simplifier used by the C++ version of FIRE with the use of the gateToFermat library by M.\ Tentukov.
In this work, several more simplifiers are connected to FIRE for the first time, through the standalone C++ library FUEL which provides access to the simplifiers.

FIRE can run both on desktop computers and on specialized nodes with 32 or more computing cores and more than 1.5 TB of RAM, on 64-bit versions of the Linux operating system. Program running time and the required amount of RAM depend on the complexity of the task, and the running time can be up to several months for real-world tasks, of which up to 95\% can be spent exclusively on the simplification of rational function coefficients when solving linear systems. In this regard, it is important to find the best programs for simplifying the coefficients, which would allow us to optimize this part of FIRE's performance.

There are many other performance considerations relevant for IBP reduction computations with the Laporta algorithm while keeping analytic dependence on kinematic and spacetime dimension variables. Such considerations include e.g.\ the ordering of integrals and ordering of equations \cite{Maierhofer:2017gsa, Bendle:2019csk}, selection of IBP identities and Lorentz-invariance identities \cite{Lee:2008tj, Smirnov:2013dia}, the use of reduction rules with abstract propagator powers (see e.g.~\cite{LiteRed, Ruijl:2017cxj}), the choice of master integral bases that avoid spurious singularities \cite{Smirnov:2020quc, Usovitsch:2020jrk}, block triangular form \cite{Liu:2018dmc, Guan:2019bcx}, syzygy equations \cite{Gluza:2010ws, Schabinger:2011dz, Larsen:2015ped, Bohm:2018bdy, Bendle:2019csk} and the related numerical unitarity method \cite{Ita:2015tya, Abreu:2017xsl, Abreu:2017hqn}. In this work, however, we focus exclusively on the simplification of rational functions in the process of solving linear systems.

The FUEL library from this work is available from the following git repository: 
\url{https://bitbucket.org/feynmanIntegrals/fuel/src/main/}

\section{Problem statement}
The purpose of this work is to select and test existing third-party programs for simplifying rational functions, and develop a C++ library FUEL for accessing the simplification functionality. The third-party programs under consideration should be able to simplify complicated expressions and be compatible with the Linux operating system. The programs must be tested for the correctness of simplification.

In order to achieve the goal, it is necessary to solve the following tasks:
\begin{itemize}
  \item Find programs, or \textit{simplifiers}, that meet the requirements, and write the FUEL library for accessing the simplifiers from C++.
  \item Test and compare the simplifiers in terms of rational function simplification performance, and select the best ones.
  \item Connect FIRE with these simplifiers via FUEL to perform IBP reduction computations, and check if they work correctly.
\end{itemize}

Caveat: our results \textbf{should not be} interpreted as a performance comparison between computer algebra systems for polynomial-oriented tasks. Rather, we test the computer algebra systems for the overall performance for tasks similar to IBP reduction when interfacing with FIRE. In particular, efficient bi-directional transfer, i.e.\ \textbf{parsing, printing, and transferring} of expressions in a text format is often an important performance bottleneck, and certain programs can be uncompetitive even when their the inherent simplification speed is excellent.

\section{Simplifiers, input data, and connecting to FIRE}

\subsection{Overview}
In general, when solving a linear system with polynomial coefficients, e.g.\ by Gaussian elimination, an intermediate coefficient to be simplified is a sum of fractions, whose numerator and denominator can also contain fractions which, in turn, contain polynomials in their numerators and denominators, as written schematically in Eq.~\eqref{4:coefficient}.
\begin{equation}\label{4:coefficient}
\sum_{k}^{}
\frac{\frac{\textrm{Poly}_{k,1}}{\textrm{Poly}_{k,2}}\cdot\frac{\textrm{Poly}_{k,3}}{\textrm{Poly}_{k,4}}}{\frac{\textrm{Poly}_{k,5}}{\textrm{Poly}_{k,6}}}
,\ \textrm{Poly}_{k,j}=\sum_{i} q_{i} \cdot x_{1}^{n_{i1}} \cdot x_{2}^{n_{i2}} \cdot \ldots \cdot x_{m}^{n_{im}},\ q_{i} \in \mathbb{Q},\ x_{h} \in X,\ n_{ih} \in \mathbb{Z}_{0}^{+} \, .
\end{equation}
A simplification is considered successful if the result is a polynomial or a fraction where the numerator and denominator are polynomials without a nontrivial GCD. The polynomials must be in either the expanded form or some nested form such as the Horner form. We do not considered other forms such as the factorized forms or partial-fractioned forms for polynomials and rational functions for most of this study.\footnote{An exception is Wolfram Mathematica, whose {\tt Together} command may choose to keep part of the expression in factorized form.} There are three additional characteristics of the problem arising from IBP reduction by FIRE. First, rational numbers can be as large as desired, that is, the simplifier must support arbitrary-precision integer and rational number arithmetic. Second, the set of possible variables is known in advance (which are kinematic variables and the spacetime dimension variable). Third, the maximum level of nested fractions does not exceed three. The second and third points are important for simplifiers that require these parameters to be passed in advance before actual calculations.

\subsection{Connecting with the simplifier}
\subsubsection{Method 1: pipe communications}
\label{subsubsec:pipe}
In order to connect the simplifier, the \textit{fork-exec} technique, popular on Unix systems, is used by FUEL to first create a copy of the current process, then run a new executable file in the context of the newly created process. If the simplifier is an executable file or comes with source code from which an executable file can be built, there is no need to do anything apart from downloading or compiling the simplifier beforing calling \textit{exec}. If the simplifier is a library, we write wrapper code to use this library and then compile the code into an executable. This technique is as universal as possible, that is, it is applicable to almost any program written in any programming language.

Communication with the spawned process is done through two specially created \textit{pipes}, the first of which is used by the parent program (e.g.\ FIRE) to write messages to the simplifier, and the second of which is used to send messages in the opposite direction, both in the text format. These techniques make it possible to create a single universal interface for the simplifiers and separate their code from the main program which creates the rational function coefficients (e.g.\ from creating linear systems of IBP equations). The final procedure for connecting the new simplifier consists of the following steps:
\begin{enumerate}
  \item Determine which command-line arguments should be passed when \textit{exec} is called.
  \item Communicate the rational function to be simplified to the simplifier in an appropriate syntax understood by the simplifier. This triggers the simplifier to parse the expression into an internal representation, simplify the expression, and prints out the result which is then sent back as the response.
  \item Write the code for parsing the simplifier response.
  \item Determine which commands to pass to the simplifier process in order for it to terminate correctly.
\end{enumerate}
The second and third items are the most time-consuming to implement, because they are unique to each of the pluggable simplifiers, and it was necessary for us to study their documentation, examples, and sometimes advice on the Internet.

\subsubsection{Method 2: library interface}
Some of the simplifiers are alternatively exposed as C++ libraries to be used directly by FUEL without pipe communications with a separate executable. In this mode of operation, every input expression is supplied as a string to the relevant library function, and a simplified expression represented as a string is returned. Parallel evaluation is possible when the relevant library is thread-safe. Other than these changes, most of the considerations of Section \ref{subsubsec:pipe} still apply. Noticeable performance gains are observed by avoiding pipe communications, as will be seen later in the paper.\footnote{One could also consider exchanging expressions as C++ objects of opaque types defined by the library, to further avoid the overhead of conversion from / to strings. However, this would not be compatible with the current version of FIRE which relies on database storage that accepts strings as data.}

\subsection{Connected simplifiers}
This is the complete list simplifiers (and the languages they are written in) which can be accessed by the current version of FUEL: CoCoA (C++), Fermat (C), FORM (C), GiNaC (C++), Macaulay2 (C/C++, Macaulay2), Maple (C, Java, Maple), Maxima (Common Lisp), Nemo (C, Julia), PARI (C), Wolfram Mathematica (C/C++, Java, Wolfram Language). This list contains both open-source and proprietary software solutions.
The issue of software licensing is an important concern, because it may affect, for example, the license under which a derived program can be distributed, the right to modify the code, and the right to distribute the code to third parties. Though we initially searched exclusively among open-source programs and libraries, we could not omit widely used proprietary computer algebra systems such as Maple and Wolfram Mathematica.

All simplifiers can be accessed by pipe communications, in which case parallel evaluation is always possible by running multiple processes. Meanwhile, CoCoA and GiNaC, and Symbolica can be accessed as C++ libraries. Among the three, only Symbolica's library is thread-safe and supports parallel evaluation.

A brief introduction to each simplifier is given in Appendix \ref{sec:simplifierIntros}.

\section{Benchmark tests with small to moderately large expressions}
\label{sec:tests}

\subsection{Testing method}
  The simplifiers under consideration are used as sequential programs when accessed via pipe communications, i.e.\ the simplifier process does not process more than one rational function expressions simultaneously. If parallel evaluation is desired, the user should spawn the required number of simplifier processes and organize parallel sending and receiving of rational function expressions.\footnote{For example, FIRE has one main execution thread and several additional threads (named FLAME) that communicate with simplifiers.} For simplifiers accessed as C++ libraries, parallel evaluation is only possible when the library is thread-safe (which is the case for Symbolica, but not CoCoA or GiNaC).

In order to test how fast the simplifiers work, a special benchmark program was written. It reads rational function expressions from a data file into memory, then with the help of a pre-selected simplifier, processes them individually: sending each expression to the simplifier, waiting for a response, receiving and parsing the response, and finally printing out the total time taken to simplify all expressions from the file.

The initialization of the simplifier (creating a process, loading its context and, for some simplifiers, passing some configuration parameters) is done once at the very beginning and takes at most several seconds. Then the process can in principle run for many hours, so the initialization cost of the simplifier is not included in the benchmark results.

Information about memory usage by the simplifier process was collected by the utility program in Ref.~\cite{Memory_measurement}, which monitors the memory usage every half a second.

The main machine used for testing has the AMD Ryzen 7 3750H CPU with base frequency 2.3 GHz, boost frequency 4.0 GHz, and 24 GB of RAM.

\subsection{Description of test data}
  Testing was performed on three sets of input data. The rational function expressions from the first set have one variable, and the rational function expressions from the second set have three variables. The number of variables is important as it can significantly affect the running time of some simplifiers. While the first two sets consist of small expressions, the third set consists of moderately large expressions; the expression size is also an important parameter that directly affects running time. The Table \eqref{5:params_of_coef_sets} shows the main quantitative characteristics of the sets used: the minimum, maximum, and average length of expressions, and the number of expressions in the file.
\begin{table}[h]
  \centering
  \begin{tabular}{|c|c|c|c|c|c|}
    \hline
    No. & No.\ of variables & Min.\ expr.\ length & Max.\ expr.\ length & Avg.\ expr.\ length & Number of coefs. \\ \hline
    1 & 1 & 1 & 5'341 & \(\sim\)29 & 692'584 \\ \hline
    2 & 3 & 1 & 2'133 & \(\sim\)33 & 971'330 \\ \hline
    3 & 2 & 232'971 & 465'943 & \(\sim\)310'628 & 12 \\ \hline
  \end{tabular}
  \caption{Parameters of sets with rational function expressions, on which testing was conducted.}
  \label{5:params_of_coef_sets}
\end{table}

An even larger expression will be considered eventually, but in a very different setting in Section \ref{sec:extraTest}.

\subsection{Results for running time}

The results obtained for the running times of the simplifiers, \emph{as accessed from FUEL}, on different test sets on the two machines are presented in Table \ref{5:time}. All values are in seconds and are rounded to one decimal place. We stress again that the running time includes not only the simplification itself, but also parsing, printing, and possibly pipe communications, so the results \textbf{should not be} considered as indicators of the inherent quality of the tested simplifiers, especially when the simplifiers are used in workflows different from ours.

\begin{table}[h]
  \centering
  \begin{tabular}{|c|c|c|c|}
    \hline
    \diagbox{Simplifier}{Set.\ No.} & 1 & 2 & 3 \\ \hline
    CoCoA & \gradient{70.6}{7}{35}{72}
          & \gradient{127.5}{13}{65}{130}
          & \gradient{84.5}{1}{5}{10} \\ \hline
    CoCoA (lib) & \gradient{56.2}{7}{35}{72} 
         & \gradient{100.3}{13}{65}{130}
         & \gradient{83.3}{1}{5}{10} \\ \hline
    Fermat & \gradient{13.9}{7}{35}{72} 
           & \gradient{15.6}{13}{65}{130}
           & \gradient{1.8}{1}{5}{10}  \\ \hline
    FORM & \gradient{62.0}{7}{35}{72} 
         & \gradient{107.2}{13}{65}{130} 
         & \gradient{1967.8}{1}{5}{10} \\ \hline
    GiNaC & \gradient{21.7}{7}{35}{72} 
          & \gradient{43.3}{13}{65}{130}
          & \gradient{4.6}{1}{5}{10} \\ \hline
    GiNaC (lib) & \gradient{11.9}{7}{35}{72} 
         & \gradient{27.2}{13}{65}{130}
         & \gradient{4.5}{1}{5}{10} \\ \hline
    Macaulay2 & \gradient{81.9}{7}{35}{72} 
             & \gradient{244.7}{13}{65}{130}
             & -   \\ \hline
    Maple & \gradient{192.1}{7}{35}{72} 
          & \gradient{276.0}{13}{65}{130} 
          & \gradient{85.4}{1}{5}{10}  \\ \hline
    Maxima & \gradient{106.6}{7}{35}{72}  
           & \gradient{182.8}{13}{65}{130} 
           & \gradient{10.9}{1}{5}{10}  \\ \hline
    Nemo & \gradient{19.6}{7}{35}{72} 
         & \gradient{35.5}{13}{65}{130} 
         & \gradient{3.4}{1}{5}{10}  \\ \hline
    PARI / GP & \gradient{10.3}{7}{35}{72} 
         & \gradient{18.3}{13}{65}{130}
         & \gradient{736.2}{1}{5}{10} \\ \hline
    Symbolica & \gradient{10.3}{7}{35}{72} 
         & \gradient{16.5}{13}{65}{130}
         & \gradient{1.8}{1}{5}{10} \\ \hline
    Symbolica (lib) & \gradient{1.9}{7}{35}{72} 
         & \gradient{4.2}{13}{65}{130}
         & \gradient{1.8}{1}{5}{10} \\ \hline
    \mythead{Wolfram\\Mathematica} & \gradient{349.4}{7}{35}{72} 
                                 & \gradient{882.0}{13}{65}{130} 
                                 & \gradient{44.2}{1}{5}{10} \\ \hline
  \end{tabular}
  \caption{Simplifier running times (in seconds) for each of the machines for three sets of input data. A simplifer is accessed via pipe communications unless the library mode is used as indicated by ``(lib)''.}
  \label{5:time}
\end{table}

We have checked that on different machines, though the absolute running times are different, the relative performance of different simplifiers remains largely unchanged.

Within each set of expressions, the simplifiers are divided into three groups, from the ones with the best performance to the ones with the worst performance: the first group includes those whose running times differ from the minimum on a given set by no more than five times, the second group includes those whose times differ by no more than 10 times, and the third group includes the rest. For clarity, the cells of the Table \ref{5:time} are colored according to this division: the first group in green, the second in yellow, and the third in red.

Let us now consider each of the test sets in more detail:

\begin{enumerate}
\item The first set is characterized by the fact that there is not more than one variable in each of the expressions. Based on performance in simplifying expressions from this test set, The first group includes Fermat, GiNaC, Nemo, PARI / GP, and Symbolica, the second group includes CoCoA and FORM, and the third group includes all others: Macaulay2, Maple, Maxima, and Wolfram Mathematica. Notice that for CoCoA, GiNaC, the library modes run faster but fall into the same groups as their pipe versions.
\item The second set differs from the previous one in that the expressions can now contain up to three variables. As can be seen from the results, the distribution across groups has not changed relative to the distribution for the first set.
\item The third set of expressions is very different from the previous two sets, in that the average expression length has increased by a factor of about 10000, although there is a smaller total number of expressions to keep the total running time manageable. We comment on the performance of a few simplifiers: CoCoA has moved from the second group to the third, showing a slight deterioration. FORM and PARI / GP have the poorest performance (as usual, with the caveat that only in the workflow considered here). While in the previous sets FORM was about 20 to 30 times worse than the best simplifier, and PARI / GP was about 5 times worse than the best one, now FORM is worse than the best-performing simplifier by about 1100 times, and PARI / GP by about 410 times. It turns out that for these simplifiers, the expression length is a defining characteristic and with its growth their speed rapidly degrades. We have not been able to test Macaulay2 on this set due to technical problems with pipe communications. For the third set, the division into groups is as follows: the first includes Fermat, GiNaC, Nemo, and Symbolica, the second includes Maxima, and the third includes CoCoA, FORM, Macaulay2, Maple, PARI / GP, and Wolfram Mathematica.
\end{enumerate}
To draw a conclusion based on these time measurements for simplifying small to moderately large expressions, we apply the following heuristic: if a simplifier is in the first group (i.e.\ the fastest group) for all the three test sets, then we will call it ``best''. If it is in the third group, that is, in the group with the slowest, then we will call it ``bad''. If neither is the case, we call it ``good''. It is important to stipulate that these labels should be understood only in conjunction with the phrase ``for this class of tasks''. According to the results of testing, the best simplifiers are Fermat, GiNaC, Nemo, and Symbolica, the group of good ones is empty, and the bad ones are CoCoA, FORM, Macaulay2, Maple, Maxima, PARI / GP, and Wolfram Mathematica. When choosing a program or library for simplifying polynomial expressions, e.g.\ for IBP reduction computations with FIRE, you should first choose from the ``best'' programs; if for some reason none of them suits you, only then consider the others. Furthermore, as mentioned previously, while the library mode (when available) of a simplifier always runs faster than its pipe mode in our single-threaded benchmark, only Symbolica supports parallel evaluation when used in the library mode.

It is also noted that two proprietary programs, Maple and Wolfram Mathematica, are grouped with the slowest simplifiers for all the three test sets. This suggests that they are unlikely to be good candidates for use as simplifiers in FIRE runs. However, it would be misleading to conclude that these two programs are inherently poor in simplifying rational functions in general, as they perform well in a different benchmark problem in Section \ref{sec:extraTest} which mainly measures inherent simplification time with less overhead in other tasks like parsing text.

Note that for challenging IBP problems, significantly larger expressions can be encountered. Therefore, in the next section, we will supplement the picture we have by additional tests involving a huge expression, while setting the tests in a different manner to shed light on computational overhead unrelated to the simplification itself.

\section{Additional tests with huge expression with low parsing overhead}
\label{sec:extraTest}
The CPU time consumed by a simplifier process consists of three parts:
\begin{enumerate}
\item Parsing the mathematical expression in a text format passed from an external program such as FIRE.
\item Simplifying the expression.
\item Printing out the expression.
\end{enumerate}
The ``inherent performance'' of the simplifier is the measured by the time spend on part (2) above, but this may not be the performance bottleneck depending on usage pattern. For example, part (1), the parsing process, can often be a bottleneck, considering that an IBP reduction run with FIRE can involve hundreds of thousands of expressions sent to the simplifier, some of which could be rather small and not inherently difficult to simplify. We will not carry out a full investigation of this issue. However, to shine some light on the impact of parsing performance, we present results from another set of test data, where the task is simplifying a single expression in 6 variables,
\begin{equation}
  \frac{(a+b+c+d+f+g)^{14}+3} {(2a+b+c+d+f+g)^{14}+4} -
  \frac{(3a+b+c+d+f+g)^{14}+5} {(4a+b+c+d+f+g)^{14}+6} \, .
\end{equation}
The expression is given in the test data file as the following line:
{\footnotesize
\begin{verbatim}
((a+b+c+d+f+g)^14+3)/((2*a+b+c+d+f+g)^14+4)-((3*a+b+c+d+f+g)^14+5)/((4*a+b+c+d+f+g)^14+6)
\end{verbatim}
}
The time taken to parse this short expression is negligible, but simplification of this expression, involving e.g.\ expanding polynomials and finding polynomial GCDs, is computationally demanding due to high powers of sub-expressions.\footnote{Note that such expressions do not arise from FIRE: even though high-degree expressions can be generated when solving IBP linear systems, the intermediate expressions are always simplified so that polynomials are in an expanded form or a nested form, and therefore there will be no explicit appearances of a single expression raised to a high power.} Not all simplifiers finish the test in reasonable time; for those that do finish, the running times are in Table \ref{tab:extraTestTime}
\begin{table}[h]
    \centering
    \begin{tabular}{|c|c|c|}
      \hline
      Simplifier & Time (seconds) \\ \hline
      Symbolica$^*$ (lib) & 5.2 \\ \hline
      Nemo & 6.9 \\ \hline
      Maple & 7.9 \\ \hline
      Fermat & 98.3 \\ \hline
      Maxima & 112.8 \\ \hline
      \mythead{Wolfram\\Mathematica} & 168.8 \\ \hline
      CoCoA$^*$ (lib) & 365 \\ \hline
    \end{tabular}
    \caption{Time taken, in seconds, by various simplifier to run the test of this section. The numbers are rounded to the nearest integer, or one decimal place if it is less than 10. Only 6 simplifiers are shown in the table. The remaining ones, when accessed from FUEL, are not able to finish the test within 1200 seconds. $^*$For Symbolica and CoCoA, the pipe mode has a similar performance as the library mode for this test since the parsing overhead is low.}
    \label{tab:extraTestTime}
\end{table}
This test is drastically different from the tests in Section \ref{sec:tests} since it artificially involves negligible parsing overhead, while the simplification itself is very demanding. The test still includes the time taken for the simplifiers to print out the results to be read by the simplifier, but printing usually has a smaller CPU footprint than parsing when large expressions are involved.\footnote{For Nemo, we found that the running time is reduced by only about 15\% when the printing of the result is turned off. For Mathematica and Maple, we found that the time used in printing is completely overwhelmed by the rest of the running time.} The results are also very different from those in Section \ref{sec:tests}. For example, Maple is now among the most performant programs in this test, either because it suffered from significant parsing overhead in previous tests or because Maple may have a relative advantage in simplifying very large rational expressions.

\section{Memory usage}
An important statistic is the maximum memory usage, which determines whether or not the \textit{out-of-memory killer} of the operating system will terminate the simplifier. This is especially a concern when simplifying complicated expressions.
We compare the maximum memory usage for the simplest test set 1 in Section \ref{sec:tests} and the most complicated expression in Section \ref{sec:extraTest}. The results are given in Table \ref{5:memory}, only for simplifers that can finish the latter test in reasonable time.\footnote{For the convenience of measurement,  we have also restricted our attention to simplifers that can run as separate proccesses accessed via pipes.} All values are rounded to integers.

\begin{table}[h]
  \centering
  \begin{tabular}{|c||c|c|}
    \hline
    Stat. & \multicolumn{2}{c|}{Max} \\ \hline
    \backslashbox{Simplifier}{Set}& Small & Huge \\ \hline
    CoCoA & \gradient{3}{1}{100}{500}
          & \gradient{242}{1}{100}{500} \\ \hline
    Fermat & \gradient{21}{1}{100}{500}
           & \gradient{22}{1}{100}{500} \\ \hline
    Maple & \gradient{20}{1}{100}{500}
          & \gradient{34}{1}{100}{500} \\ \hline
    Maxima & \gradient{931}{1}{100}{500}
           & \gradient{953}{1}{100}{500} \\ \hline
    Nemo & \gradient{409}{1}{100}{500}
         & \gradient{410}{1}{100}{500} \\ \hline
    Symbolica & \gradient{2}{1}{100}{500}
         & \gradient{20}{1}{100}{500} \\ \hline
    \mythead{Wolfram\\Mathematica} & \gradient{121}{1}{100}{500}
                                 & \gradient{225}{1}{100}{500} \\ \hline
  \end{tabular}
  \caption{The maximum amount of used RAM by simplifiers when accessed via FUEL, in megabytes, for three sets of input expressions. The ``small'' test set refers to set 1 in Section \ref{sec:tests}, and the ``huge'' test set refers to the expression in Section \ref{sec:extraTest}.}
  \label{5:memory}
\end{table}


Except for CoCoA and Wolfram Mathematica, the simplifiers in the table exhibit memory usage that depends rather mildly on the complexity of the expressions. One should keep in mind that even standard laptops now have at least 8GB of RAM and server systems can have several hundred GB of memory. For IBP reduction with FIRE, the memory consumption from storing a large number of reduction rules (in which the rational functions are already simplified) can be a much more serious concern.

\section{Tests with FIRE}
\label{sec:fireTest}
\subsection{A simple example}
\begin{figure}
  \centering
  \includegraphics[width=.36\textwidth]{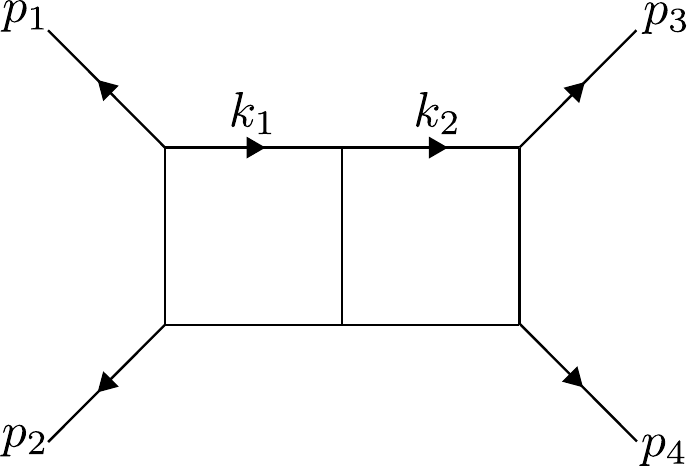}
  \caption{The double box diagram to be tested for IBP reduction in FIRE6 calling various different simplifiers.}
  \label{fig:doublebox}
\end{figure}

We run the double box IBP reduction example in FIRE6. This IBP problem is very simple by current standards and should be considered as a preliminary test, as the main focus of this work is presenting FUEL and standalone benchmark tests. The double box diagram is shown in Fig.~\ref{fig:doublebox}.
We reduce a rank-2 tensor integral with numerator
$$(k_2+p_1)^2 (k_1-p_3)^2$$
using only one worker thread. The statistics printed out at the end of FIRE runs are in Table \ref{tab:fireTest}. FIRE runs involve two stages, forward elimination and backward substitution. While back substitution consumes a small percentage of the total running time in this simple problem, it can become the dominant part in more complicated IBP problems. Therefore we have shown the ``substitution time'' as a separate column in the table.
\begin{table}[h]
    \centering
    \begin{tabular}{|c|c|c|c|}
      \hline
      Simplifier & Total Time & Substitution time & Memory usage (FIRE + simplifier) \\ \hline
      Symbolica (lib) & 14.6 & 1.1 & 15.6 \\ \hline
      Fermat & 19.8 & 1.8 & 23.0 \\ \hline
      Pari / GP & 21.6 & 2.3 & 14.1 \\ \hline
      Nemo & 32.8 & 2.4 & 431.7 \\ \hline
      GiNaC (lib) & 27.6 & 6.0 & 14.6 \\ \hline
      CoCoA (lib) & 73.2 & 4.7 & 16.0 \\ \hline
      FORM & 78.5 & 5.8 & 14.4 \\ \hline
      Maxima & 131.2 & 6.7 & 955.8 \\ \hline
      Macaulay & 152.5 & 12.1 & 350.9 \\ \hline
      Maple & 177.6 & 6.5 & 105.1 \\ \hline
      \mythead{Wolfram\\Mathematica} & 581.0 & 22.2 & 146.1 \\ \hline
    \end{tabular}
    \caption{Performance of various simplifier when used by FIRE to reduce a rank-2 tensor integral for the massless two-loop double box. The running times are in seconds, while the memory usage is in units of MB.}
    \label{tab:fireTest}
\end{table}
In this test, the performance of the simplifiers relative to each other is very similar to the situation in test sets 1 and 2 in Section \ref{sec:tests}. Based on the data for test 3 in Section \ref{sec:tests}, it is likely that the situation can change dramatically for highly demanding IBP reduction problems.

\subsection{A complicated example}
\begin{figure}
  \centering
  \includegraphics[width=0.36\textwidth]{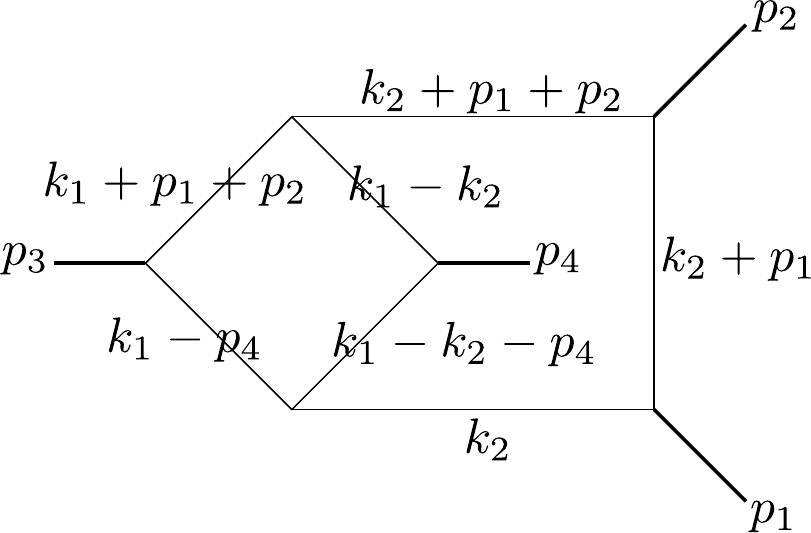}
  \caption{A nonplanar double box diagram topology, with massless internal lines and off-shell external lines.}
  \label{fig:offshell_nonplanar_dbox}
\end{figure}

We now show the application to a demanding IBP problem for nonplanar double box integrals with massless internal lines and off-shell external lines. Three of the four external lines have the same virtuality. The problem arises from a computation of off-shell three-particle form factors in $\mathcal N=4$ super-Yang-Mills theory \cite{BelitskyInPreparation}, following up on an earlier computation of two-particle form factors \cite{Belitsky:2022itf, Belitsky:2023ssv}. The diagram is shown in Fig.~\ref{fig:offshell_nonplanar_dbox}.
The kinematics is
\begin{align*}
  & p_1^2 = p_2^2 = p_3^2 = -m^2 , \\
  & (p_1 + p_2)^2 = -u , \\
  & (p_2 + p_3)^2 = -v , \\
  & (p_1 + p_3)^2 = -w \, .
\end{align*}
We perform the IBP reduction of top-level integrals with these three different numerators:
\begin{enumerate}
\item the unit numerator $1$
\item $(k_1 + p_1)^2$
\item $(k_2 - p_4)^2$
\item $(k_1 + p_1)^2 (k_2 - p_4)^2$
\end{enumerate}
There are 97 master integrals after IBP reduction. As a form of parallelization, FIRE allows the user to set all but one master integrals to zero and compute the coefficient of the one chosen master integral. For the purpose of benchmarking, we only compute the coefficient of the bottom-level ``sunrise'' master integral which contains the propagators $(k_1-p_4)^2$, $(k_1-k_2)^2$, and $(k_2+p_1)^2$. Five variables are involved in the IBP reduction, including the spacetime dimension $d$ and the kinematic variables $m$, $u$, $v$, $w$. The large number of variables makes the simplification of rational functions computationally demanding during the IBP reduction. The time needed to complete the task for Fermat, Nemo, and Symbolica are shown in Table \ref{tab:fireFiveVarTest}.\footnote{A different computer is used for this benchmark as the previous tests were carried out at an earlier time. The CPU used here is the Intel Xeon Gold 6240 Processor (24.75M Cache, 2.60 GHz). We allowed up to 16 simplifier processes to run in parallel, but due to limitations of parallelization capabilities when only a single master integral is targeted, only a single thread is active for almost the entire duration of the run, regardless of the simplifier used.}
\begin{table}[h]
    \centering
    \begin{tabular}{|c|c|c|}
      \hline
      Simplifier & Total Time ($\times$ 1000 seconds) \\ \hline
      Symbolica (lib) & 8.9 \\ \hline
      Nemo & 11.9 \\ \hline
      Fermat & 109.7 \\ \hline
    \end{tabular}
    \caption{Performance of various simplifier when used by FIRE to reduce sample integrals for the nonplanar double box with massless internal lines and off-shell external lines, involving four kinematic variables and the spacetime variable in rational function simplification.}
    \label{tab:fireFiveVarTest}
\end{table}
We can see that Symbolica and Nemo offers significant speedups compared with Fermat, as is the case in the artificial benchmark in Section \ref{sec:extraTest} outside of FIRE.

\section{Conclusions}
We have presented a new C++ library FUEL for simplifying rational function expressions, in light of ongoing efforts to improve the performance of integration-by-parts reduction for complicated Feynman integral calculations. As a standalone library, FUEL can also potentially find applications in other areas. Under a universal interface, FUEL allows a flexible choice of \textit{simplifiers}, i.e.\ existing computer algebra programs or libraries, as the underlying computation engine. FUEL grew out of FIRE's original interface to Fermat, the latter being a computer algebra system written by Robert Lewis. FUEL has two modes of operation. The first mode is the pipe mode as originally used by FIRE, based on inter-process communication over unix pipes, sending and receiving text expressions to a third-party simplifier program in a text format. The other mode is the library mode, i.e.\ directly linking with third-party libraries without spawning child processes. With both mode available, the setup allows for maximum flexibility in connecting with any simplifier written in any programming language. In the pipe mode, parallel computation is achieved by running multiple processes of the same simplifier (or even different simplifiers catering to different types of expressions, experimentally supported by FUEL). In the library mode, parallel computation is possible when the third-party library is thread-safe (which is the case only for Symbolica so far).
Good performance under the pipe mode requires the simplifier to be fast in both the key task of simplifying mathematical expressions and overhead tasks such as parsing text inputs, which makes certain simplifiers (such as recent versions of Maple) uncompetitive for our purpose even when they have a reputation for fast manipulations of polynomials and rational functions.

In the current version of FUEL, we have implemented connections with 11 different simplifiers, including CoCoA (pipe or library mode), Fermat, FORM, GiNaC (pipe or library mode), Macaulay2, Maple, Maxima, Nemo, PARI / GP, Symbolica (pipe or library mode), and Wolfram Mathematica. Nemo has been augmented by a dedicated Julia package we wrote (distributed with FUEL) to enable fast parsing of mathematical expressions. Artificial benchmark tests with small to moderately large expressions are presented in Section \ref{sec:tests}. Symbolica is the fastest simplifier (or in a tie with the fastest simplifiers) for all the test sets, while Fermat, Nemo and GiNaC are also consistently among the fastest simplifiers. Pari / GP is very fast for the first two sets of test data involving shorter expressions (which likely mimic less demanding FIRE runs), but performs very poorly in the third data set involving moderately large expressions.

An additional special-purpose test is presented in Section \ref{sec:extraTest}. Compared with the main tests discussed above, this test minimizes the overhead of text parsing but is extremely computationally intensive in the simplification itself. Here Fermat has dropped to the third place in the ranking of the fast programs, led by Symbolica, Nemo, and Maple. Both Pari / GP and GiNaC (among others) have failed to complete the test before the 20-minute timeout. We plan to explore using more than one simplifier in a single C++ program, e.g.\ FIRE, given their different performance characteristics for different problems.

A private experimental version of FIRE has been linked with FUEL to perform IBP reduction of Feynman integrals. We have first demonstrated a very simple IBP reduction example for the two-loop double box, which can be completed by FIRE with any of the 10 connected simplifiers. The time required by the run for each simplifier has been tabulated, and the results are broadly consistent with those from the simpler test sets in the artificial benchmarks of Section \ref{sec:tests}. Next, we have tested the IBP reduction performance for a non-planar double box family of integrals which involve four kinematic variables besides the spacetime variable. In this case, Symbolica and Nemo significantly outperform Fermat, while GiNaC and Pari / GP cannot finish the test in reasonable time, mirroring the demanding standalone benchmark of Section \ref{sec:extraTest}. We leave further improvements to further work, and a first application to a concrete physics problem using our software is being carried out by other authors \cite{BelitskyInPreparation}.

\section{Acknowledgments}
We thank the authors of FLINT, Nemo and FORM for help with our questions about the software in mailing lists and/or private communications.
We especially thank the author of Symbolica, Ben Ruijl, for tirelessly answering our questions and customizing the software to integrate with our library.
The work of Alexander Smirnov was supported by the Russian Science Foundation under the agreement No.\ 21-71-30003.
M.Z.’s work is supported in part by the U.K.\ Royal Society through Grant URF\textbackslash R1\textbackslash 20109. For the purpose of open access, the authors have applied a Creative Commons Attribution (CC BY) license to any Author Accepted Manuscript version arising from this submission.

\appendix

\section{Overviews of connected simplifiers}

\subsection{CoCoA}

CoCoA \cite{CoCoA} is a computer algebra system for computing in polynomial rings. The development of the system began in 1987 in Pascal, hence its Pascal-like syntax, and it was later rewritten in C, and yet a little later a C++ library, CoCoALib \cite{CoCoALib}, appeared, and the latter library was used in our work. CoCoA allows to perform calculations in rings of polynomials of many variables with rational or integer coefficients, as well as over ideals of these rings. The user can redefine the polynomial ring used as well as various homomorphisms for converting elements from one ring to another. According to CoCoA's authors, the Gröbner basis is used as the key mechanism for efficient computations in commutative algebra.

In order for CoCoA to handle a factor of the form Eq.~\eqref{4:coefficient} passed to it, we need to specify, in the C++ constructor, a field that it belongs to. To do this, we specify the appropriate \textit{ring of integers}, \textit{ring of fractions}, and \textit{ring of polynomials}, combining and substituting one into the other to get the desired field. Then it is possible to supply a string representation of the rational function and get a simplified representation from it.

The fully expanded form is used for polynomial representations in our use of CoCoA. In this paper, we use the version CoCoA 0.99715.

\subsection{Fermat}
Fermat \cite{Fermat} is a computer algebra system developed by Robert Lewis, with the goal of being fast and memory efficient, covering ``arithmetic of arbitrarily long integers and fractions, multivariate polynomials, symbolic calculations, matrices over polynomial rings, graphics, and other numerical calculations''. Fermat has influenced research in fast rational function arithmetic in computer algebra \cite{monagan2017fermat}. Until recently, Fermat was the only simplifier used by the C++ version of FIRE. Fermat is also the main simplifier in two other IBP reduction programs, Kira and Reduze.

The output of Fermat expresses polynomials in the Horner form. In this paper, we use the version Fermat 5.17.

\subsection{Form}
FORM \cite{FORM} is a computer algebra system, which the authors themselves prefer to call a system for formula conversions. It has been in development since 1984, and its original goal was to simplify calculations in quantum field theory. FORM is written in C and accepts input in a special programming language, which is then interpreted and executed. In other words, FORM is not interactive but operates as a batch-processing program. FORM's language has many features: it has an advanced preprocessor with more than sixty commands, several types of variables: symbols, vectors, indices, functions, sets, more than a hundred commands controlling the execution, output and properties of variables, and more than eighty functions. Besides the regular version of the program, there are also two parallelized versions: ParFORM, which runs on a system with independent nodes, each using its own processor, memory and disk, and TFORM, which uses POSIX threads to better expose multiprocessor capabilities on shared-memory machines. FORM is distributed under the GNU GPL license.

For some special cases it may be necessary to override the standard settings that control how FORM works, such as the maximum size of the substitution tree, the maximum size of a term that does not require additional allocations, the size of the I/O buffers, and other settings. In order to simplify a large expression, we need to redefine several settings, otherwise the program would stop due to insufficient memory.

The fully expanded form is used for polynomial representations in our use of FORM. In this paper, we use the version FORM 4.2.1.

\subsection{GiNaC}
GiNaC \cite{GiNaC} is a C++ library for computer algebra, initially designed for Feynman diagram calculations. Contrary to many other computer algebra systems which come with their own proprietary interactive languages, GiNaC emphasizes programmatic use, extensibility and interoperability with other programs within a statically-typed compiled language (C++). Besides features commonly found in most systems, such as big integers and polynomial simplification, GiNaC offers functionalities useful for Feynman diagram calculations, such as handling of expressions involving Lorentz, Dirac, and color indices. In high energy physics research, GiNaC is perhaps most well known for its support for numerical evaluations of special functions known as multiple polylogarithms. A fork of GiNaC, PyNaC \cite{pynac}, was used as a core component of SageMath \cite{zimmermann2018computational}, a flagship open-source computer algebra system.

The fully expanded form is used for polynomial representations in our use of GiNaC. In this paper, we use the version GiNaC 1.8.2.

\subsection{Macaulay2}
Macaulay2 \cite{Macaulay2} is system for computation in algebraic geometry and commutative algebra, covering functionalities such as Groebner bases, free resolutions of modules, Betti numbers, primary decomposition of ideals, etc. Macaulay2 has been used in research in applying computational algebraic geometry to IBP reduction \cite{Zhang:2016kfo}.

The fully expanded form is used for polynomial representation in our use of Macaulay2. In this paper, we use the Macaulay2 version 1.19.1+ds-6.

\subsection{Maple}
Maple is a general-purpose computer algebra system. It is widely used in high energy physics. For example, the IBP reduction program AIR \cite{AIR} is written in Maple. Compared with its competitor Wolfram Mathematica, Maple offers a more conventional ALGOL/C-like programming language. Maple has seen continuous and active developments in high-performance algorithms relevant for polynomials and rational functions (see e.g.~\cite{monagan2013poly, monagan2017fermat, monagan2022speeding}). We use the \textit{normal} function in Maple to simplify rational functions, with the option \textit{expanded} to prevent polynomials from being kept in factorized forms. Therefore the fully expanded form is used for polynomial representation in our use of Maple.

In this paper, we use the version Maple 2022.

\subsection{Maxima}
Maxima \cite{Maxima} is a computer algebra system, a descendant of Macsyma, which allows many different operations on symbolic and numeric expressions. Maxima is self-described as a ``fairly complete computer algebra system'' and can be used to e.g.\ differentiate, integrate, solve Laplace transforms, and construct graphs. Maxima, like its ancestor, is written in Lisp. It is distributed under the GNU GPL license. SageMath \cite{zimmermann2018computational} uses Maxima internally for certain nontrivial computations such as symbolic integration and taking limits.

Maxima has rich functionality for simplifications; it offers several functions for the user to choose from: \textit{rat}, \textit{ratsimp}, \textit{fullratsimp}, \textit{radscan} and many flags that affect how the functions work. Some functions are mainly intended to simplify rational expressions, while others can simplify expressions containing logarithmic, exponential and power functions. Some perform simplification once over the expression, while others do it until the resulting expression stops to change.

In addition, several flags have been included to make it easier to parse the rational functions simplified by Maxima: \textit{display2d:false} disables 2D output, \textit{stardisp:true} removes unnecessary multiplication signs, and \textit{nolabels:true} allows to remove unnecessary I/O labels for entered and resulting expressions.

A partial expanded form is used for polynomial representation in our use of Maxima. In this paper, we use the version Maxima 5.45.1.

\subsection{Nemo (with a custom parser and printer)}

Nemo \cite{Nemo} is a computer algebra system for the Julia programming language \cite{bezanson2017julia}, and it aims to ``provide highly performant commutative algebra, number theory, group theory and discrete geometry routines.'' It provides a fast Julia interface to C/C++ libraries such as FLINT \cite{hart2011flint}. FLINT provides efficient operations for polynomials over a variety of number fields such as rational numbers and prime fields. Benefiting from the EU-funded OpenDreamKit project for open-source computer algebra, FLINT gained fast code for multivariate polynomials. Meanwhile, the Julia code in Nemo provides, among other functionalities, operations for rational functions that build upon polynomial operations of FLINT. In the current version of FUEL, we always use Nemo's sparse multivariate polynomials and associated rational functions. In the univariate case, specialized routines from Nemo and FLINT can be faster but have not been used in our work due to a lack of implementation effort on our side.

A previous internal version of FUEL calls Nemo from a Julia REPL session (i.e.\ an interactive user session), and the performance was poor due to the overhead of parsing. The parser of the Julia REPL is designed to process arbitrary syntax in the Julia language and is relatively slow for our special purpose of parsing rational function expressions. Fortunately, taking advantage of Julia's JIT compilation, we are able to write a fast parser for mathematical expressions based on a variation of the well-known shunting-yard algorithm. The parser is included in a Julia package, \textit{RationalCalculator}, bundled with FUEL. Additionally, the aforementioned package supports printing out calculation results in a format that is not human-readable but instead optimized for transfer of expressions between the simplifier and FUEL. Human readable output can be re-enabled by calling a routine in FUEL, e.g.\ when FIRE writes out IBP reduction tables, so that the final IBP reduction table from the FIRE run is unaffected.

The fully expanded form is used for polynomial representation in our use of Nemo. In this paper, we use the versions Nemo 0.33.1 and Julia 1.8.5.

\subsection{PARI / GP}
PARI / GP is a computer algebra system focused on number theory, developed at the University of Bordeaux. It can compute ``factorizations, algebraic number theory, elliptic curves, modular forms, L functions'', etc.~\cite{PARI2}. PARI is a C library, while GP is the front-end that allows interactive use. We use GP as the expression simplifier.

A Horner-like form is used for polynomial representation by PARI / GP. In this paper, we use the version GP 2.13.3.

\subsection{Symbolica}
Symbolica is a new computer algebra system being developed by Ben Ruijl \cite{Symbolica}, which aims to be a modernized incarnation of FORM \cite{FORM}, e.g.\ by using iterators in modern programming languages to express the stream processing of large expressions that cannot be packed into RAM. Symbolica is written in Rust, with interfaces for C/C++ and Python. Though at an early stage of development, Symbolica already contains highly optimized code for polynomial GCD and rational function arithmetic.

The fully expanded form is used for polynomial representation in our use of Symbolica. In this paper, we use the version a8f72f (the commit hash) from the GitHub repository \url{https://github.com/benruijl/symbolica.git}.

\subsection{Wolfram Mathematica}
Wolfram Mathematica is the most widely used general-purpose computer algebra system as of today, at least in theoretical high energy physics research. It was initially developed by Stephen Wolfram, with influences from Maxima and Wolfram's earlier system, SMP. Mathematica offers a high-level language emphasizing functional programming and term rewriting (called \emph{Replacement} in Mathematica). As of today, Mathematica encompasses a huge range of functionalities in symbolic and numerical computing. Many software packages and research data in high energy physics are published in Mathematica's formats. FIRE, even when used in the C++ mode for the main computation, uses Mathematica for pre-processing user-supplied integral family information and various post-processing tasks such as loading reduction tables and finding symmetry rules relating master integrals. We use the \textit{Together} function in Mathematica to simplify rational functions, and the output may consider either factorized or expanded polynomials.

In this paper, we use the version Mathematica 13.0.

\clearpage


\begin{thebibliography}{100}

\bibitem{algorithm_to_calculete_feynman_integral}
K.~G. Chetyrkin and F.~V. Tkachov.
\newblock {Integration by Parts: The Algorithm to Calculate beta Functions in 4
  Loops}.
\newblock {\em Nucl. Phys. B}, 192:159--204, 1981.

\bibitem{Laporta:2000dsw}
S.~Laporta.
\newblock {High precision calculation of multiloop Feynman integrals by
  difference equations}.
\newblock {\em Int. J. Mod. Phys. A}, 15:5087--5159, 2000.

\bibitem{CoCoA}
J.~Abbott, A.~M. Bigatti, and L.~Robbiano.
\newblock {CoCoA}: a system for doing {C}omputations in {C}ommutative
  {A}lgebra.
\newblock \url{http://cocoa.dima.unige.it}.

\bibitem{CoCoALib}
J.~Abbott and A.~M. Bigatti.
\newblock {CoCoALib}: a c++ library for doing {C}omputations in {C}ommutative
  {A}lgebra.
\newblock Available at \texttt{http://cocoa.dima.unige.it/cocoalib}.

\bibitem{Fermat}
R.~H. Lewis.
\newblock {Fermat: A Computer Algebra System for Polynomial and Matrix
  Computation}.
\newblock \url{http://home.bway.net/lewis/}.
\newblock Accessed: 2023-02-18.

\bibitem{FORM}
J.~A.~M. Vermaseren.
\newblock \url{https://www.nikhef.nl/~form/maindir/maindir.html}.

\bibitem{GiNaC}
\url{https://ginac.de/}.

\bibitem{Macaulay2}
{Grayson, Daniel R. and Stillman, Michael E.}
\newblock {Macaulay2, a software system for research in algebraic geometry}.
\newblock \url{http://www.math.uiuc.edu/Macaulay2/}.

\bibitem{Maple}
{Maplesoft, a division of Waterloo Maple Inc.}
\newblock {The essential tool for mathematics}.
\newblock \url{https://www.maplesoft.com/products/Maple/}.

\bibitem{Maxima}
{Maxima, a Computer Algebra System. Version 5.43.2}.
\newblock \url{https://maxima.sourceforge.io/}.

\bibitem{Nemo}
Claus Fieker, William Hart, Tommy Hofmann, and Fredrik Johansson.
\newblock Nemo/hecke: computer algebra and number theory packages for the julia
  programming language.
\newblock In {\em Proceedings of the 2017 acm on international symposium on
  symbolic and algebraic computation}, pages 157--164, 2017.

\bibitem{PARI2}
The~PARI Group.
\newblock {PARI/GP version \texttt{2.11.2}, Univ. Bordeaux, 2022}.
\newblock \url{http://pari.math.u-bordeaux.fr/}.

\bibitem{Symbolica}
Ben Ruijl.
\newblock {Symbolica}.
\newblock \url{https://symbolica.io/}.
\newblock Accessed: 2023-06-23.

\bibitem{Math}
{Wolfram Research, Inc.}
\newblock {Mathematica, Version 13.1}.
\newblock \url{https://www.wolfram.com/mathematica}.

\bibitem{Smirnov:2013dia}
A.~V. Smirnov and V.~A. Smirnov.
\newblock {FIRE4, LiteRed and accompanying tools to solve integration by parts
  relations}.
\newblock {\em Comput. Phys. Commun.}, 184:2820--2827, 2013.

\bibitem{Smirnov:2014hma}
Alexander~V. Smirnov.
\newblock {FIRE5: a C++ implementation of Feynman Integral REduction}.
\newblock {\em Comput. Phys. Commun.}, 189:182--191, 2015.

\bibitem{FIRE6}
A.~V. Smirnov and F.~S. Chuharev.
\newblock {FIRE6: Feynman Integral REduction with Modular Arithmetic}.
\newblock {\em Comput. Phys. Commun.}, 247:106877, 2020.

\bibitem{AIR}
Charalampos Anastasiou and Achilleas Lazopoulos.
\newblock {Automatic integral reduction for higher order perturbative
  calculations}.
\newblock {\em JHEP}, 07:046, 2004.

\bibitem{Reduze}
A.~von Manteuffel and C.~Studerus.
\newblock {Reduze 2 - Distributed Feynman Integral Reduction}.
\newblock 1 2012.

\bibitem{LiteRed}
Roman~N. Lee.
\newblock {LiteRed 1.4: a powerful tool for reduction of multiloop integrals}.
\newblock {\em J. Phys. Conf. Ser.}, 523:012059, 2014.

\bibitem{Maierhofer:2017gsa}
Philipp Maierh\"ofer, Johann Usovitsch, and Peter Uwer.
\newblock {Kira\textemdash{}A Feynman integral reduction program}.
\newblock {\em Comput. Phys. Commun.}, 230:99--112, 2018.

\bibitem{Kira}
Philipp Maierh\"ofer and Johann Usovitsch.
\newblock {Kira 1.2 Release Notes}.
\newblock 12 2018.

\bibitem{Klappert:2020nbg}
Jonas Klappert, Fabian Lange, Philipp Maierh\"ofer, and Johann Usovitsch.
\newblock {Integral reduction with Kira 2.0 and finite field methods}.
\newblock {\em Comput. Phys. Commun.}, 266:108024, 2021.

\bibitem{FIRE_usage_1}
Johannes~M. Henn, Alexander~V. Smirnov, Vladimir~A. Smirnov, and Matthias
  Steinhauser.
\newblock {A planar four-loop form factor and cusp anomalous dimension in QCD}.
\newblock {\em JHEP}, 05:066, 2016.

\bibitem{FIRE_usage_2}
Johannes Henn, Alexander~V. Smirnov, Vladimir~A. Smirnov, Matthias Steinhauser,
  and Roman~N. Lee.
\newblock {Four-loop photon quark form factor and cusp anomalous dimension in
  the large-$N_c$ limit of QCD}.
\newblock {\em JHEP}, 03:139, 2017.

\bibitem{FIRE_usage_3}
Roman~N. Lee, Alexander~V. Smirnov, Vladimir~A. Smirnov, and Matthias
  Steinhauser.
\newblock {The $n_f^2$ contributions to fermionic four-loop form factors}.
\newblock {\em Phys. Rev. D}, 96(1):014008, 2017.

\bibitem{Bendle:2019csk}
Dominik Bendle, Janko B\"ohm, Wolfram Decker, Alessandro Georgoudis,
  Franz-Josef Pfreundt, Mirko Rahn, Pascal Wasser, and Yang Zhang.
\newblock {Integration-by-parts reductions of Feynman integrals using Singular
  and GPI-Space}.
\newblock {\em JHEP}, 02:079, 2020.

\bibitem{Lee:2008tj}
R.~N. Lee.
\newblock {Group structure of the integration-by-part identities and its
  application to the reduction of multiloop integrals}.
\newblock {\em JHEP}, 07:031, 2008.

\bibitem{Ruijl:2017cxj}
B.~Ruijl, T.~Ueda, and J.~A.~M. Vermaseren.
\newblock {Forcer, a FORM program for the parametric reduction of four-loop
  massless propagator diagrams}.
\newblock {\em Comput. Phys. Commun.}, 253:107198, 2020.

\bibitem{Smirnov:2020quc}
A.~V. Smirnov and V.~A. Smirnov.
\newblock {How to choose master integrals}.
\newblock {\em Nucl. Phys. B}, 960:115213, 2020.

\bibitem{Usovitsch:2020jrk}
Johann Usovitsch.
\newblock {Factorization of denominators in integration-by-parts reductions}.
\newblock 2 2020.

\bibitem{Liu:2018dmc}
Xiao Liu and Yan-Qing Ma.
\newblock {Determining arbitrary Feynman integrals by vacuum integrals}.
\newblock {\em Phys. Rev. D}, 99(7):071501, 2019.

\bibitem{Guan:2019bcx}
Xin Guan, Xiao Liu, and Yan-Qing Ma.
\newblock {Complete reduction of integrals in two-loop five-light-parton
  scattering amplitudes}.
\newblock {\em Chin. Phys. C}, 44(9):093106, 2020.

\bibitem{Gluza:2010ws}
Janusz Gluza, Krzysztof Kajda, and David~A. Kosower.
\newblock {Towards a Basis for Planar Two-Loop Integrals}.
\newblock {\em Phys. Rev. D}, 83:045012, 2011.

\bibitem{Schabinger:2011dz}
Robert~M. Schabinger.
\newblock {A New Algorithm For The Generation Of Unitarity-Compatible
  Integration By Parts Relations}.
\newblock {\em JHEP}, 01:077, 2012.

\bibitem{Larsen:2015ped}
Kasper~J. Larsen and Yang Zhang.
\newblock {Integration-by-parts reductions from unitarity cuts and algebraic
  geometry}.
\newblock {\em Phys. Rev. D}, 93(4):041701, 2016.

\bibitem{Bohm:2018bdy}
Janko B\"ohm, Alessandro Georgoudis, Kasper~J. Larsen, Hans Sch\"onemann, and
  Yang Zhang.
\newblock {Complete integration-by-parts reductions of the non-planar
  hexagon-box via module intersections}.
\newblock {\em JHEP}, 09:024, 2018.

\bibitem{Ita:2015tya}
Harald Ita.
\newblock {Two-loop Integrand Decomposition into Master Integrals and Surface
  Terms}.
\newblock {\em Phys. Rev. D}, 94(11):116015, 2016.

\bibitem{Abreu:2017xsl}
S.~Abreu, F.~Febres~Cordero, H.~Ita, M.~Jaquier, B.~Page, and M.~Zeng.
\newblock {Two-Loop Four-Gluon Amplitudes from Numerical Unitarity}.
\newblock {\em Phys. Rev. Lett.}, 119(14):142001, 2017.

\bibitem{Abreu:2017hqn}
Samuel Abreu, Fernando Febres~Cordero, Harald Ita, Ben Page, and Mao Zeng.
\newblock {Planar Two-Loop Five-Gluon Amplitudes from Numerical Unitarity}.
\newblock {\em Phys. Rev. D}, 97(11):116014, 2018.

\bibitem{Memory_measurement}
\url{https://raw.githubusercontent.com/pixelb/ps_mem/master/ps_mem.py}.

\bibitem{BelitskyInPreparation}
A.~V. Belitsky, L.~V. Bork, and V.~A. Smirnov.
\newblock {In preparation}.

\bibitem{Belitsky:2022itf}
A.~V. Belitsky, L.~V. Bork, A.~F. Pikelner, and V.~A. Smirnov.
\newblock {Exact Off Shell Sudakov Form Factor in N=4 Supersymmetric Yang-Mills
  Theory}.
\newblock {\em Phys. Rev. Lett.}, 130(9):091605, 2023.

\bibitem{Belitsky:2023ssv}
A.~V. Belitsky, L.~V. Bork, and V.~A. Smirnov.
\newblock {Off-shell form factor in N=4 sYM at three loops}.
\newblock 6 2023.

\bibitem{monagan2017fermat}
Michael Monagan and Roman Pearce.
\newblock Fermat benchmarks for rational expressionals in maple.
\newblock {\em ACM Communications in Computer Algebra}, 50(4):188--190, 2017.

\bibitem{pynac}
\url{https://github.com/pynac/pynac}.

\bibitem{zimmermann2018computational}
Paul Zimmermann, Alexandre Casamayou, Nathann Cohen, Guillaume Connan, Thierry
  Dumont, Laurent Fousse, Fran{\c{c}}ois Maltey, Matthias Meulien, Marc
  Mezzarobba, Cl{\'e}ment Pernet, et~al.
\newblock {\em Computational mathematics with SageMath}.
\newblock SIAM, 2018.

\bibitem{Zhang:2016kfo}
Yang Zhang.
\newblock {Lecture Notes on Multi-loop Integral Reduction and Applied Algebraic
  Geometry}.
\newblock 12 2016.

\bibitem{monagan2013poly}
Michael Monagan and Roman Pearce.
\newblock Poly: A new polynomial data structure for maple 17.
\newblock {\em ACM Communications in Computer Algebra}, 46(3/4):164--167, 2013.

\bibitem{monagan2022speeding}
Michael Monagan.
\newblock Speeding up polynomial gcd, a crucial operation in maple.
\newblock {\em Maple Transactions}, 2(1), 2022.

\bibitem{bezanson2017julia}
Jeff Bezanson, Alan Edelman, Stefan Karpinski, and Viral~B Shah.
\newblock Julia: A fresh approach to numerical computing.
\newblock {\em SIAM review}, 59(1):65--98, 2017.

\bibitem{hart2011flint}
William~B Hart.
\newblock Flint: Fast library for number theory.
\newblock {\em Computeralgebra-Rundbrief: Vol. 49}, 2011.

\end{thebibliography}
\bibliographystyle{unsrt}

\end{document}